\documentclass[doublecol,figures]{epl3}

\title{Transverse momentum spectra of $D$ and $B$ mesons in hadron
collisions at high energies}

\author{G. I. Lykasov\inst{1} \and Z. M. Karpova\inst{1} 
\and M. N. Sergeenko\inst{2} \and V. A. Bednyakov\inst{1}}

\institute{\inst{1} Joint Institute for Nuclear Research -
  Dubna 141980, Moscow region, Russia\\
\inst{2} State University of Transport - 34 Kirov Street, Gomel
246653, Belarus}


\abstract{Transverse momentum spectra of charmed and beauty mesons
produced in proton-proton and proton-antiproton collisions at high
energies are analyzed within the modified quark-gluon string model
(QGSM) including the internal motion of quarks in colliding 
hadrons. It is shown that this approach can describe rather 
satisfactorily the experimental data at not large values of the transverse 
momentum where the NLO QCD calculation has a big uncertainty.
We also show that using both the QGSM and the NLO QCD one can describe
these data in a wide region of transverse momenta and give
some predictions for the future LHC experiments.} 

\begin{document}
\maketitle

\section{Introduction}
Various approaches of perturbative QCD including the
next-to-leading order calculations (NLO QCD) have been applied to
construct distributions of quarks in a proton. The theoretical
analysis of the lepton deep inelastic scattering (DIS) off protons
and nuclei provides rather realistic information on the
distribution of light quarks like $u,d,s$ in a proton. However, to
find a reliable distribution of heavy quarks like $c({\bar c})$
and especially $b({\bar b})$ in a proton describing the
experimental data on the DIS is a non-trivial task. 
It is mainly
due to small values of $D$ and $B$ meson yields in the DIS at
existing energies. Even at the Tevatron energies the $B$- meson
yield is not so large. At LHC energies the multiplicity of these
%$D$ and $B$
mesons produced in $pp$ collisions will be significantly
larger. Therefore one can try to extract a new information on the
distribution of these heavy quarks in a proton. In this paper we
suggest to study the distribution of heavy quarks like $c({\bar
c})$ and $b({\bar b})$ in a proton from the analysis of the future
LHC experimental data.

The multiple hadron production in hadron-nucleon collisions at
high energies and large transfers is usually analyzed
within the hard parton scattering model (HPSM) suggested in 
\cite{Efremov,FF}. This model was applied to
the charmed meson production both in proton-proton and meson-proton 
interactions at high energies, see for example \cite{Bednyakov:1995}. 
The HPSM is significantly improved by applying the QCD parton approach 
implemented in the modified minimal-subtraction renormalization and 
factorization scheme.
The first calculation scheme is the so-called massive scheme
or fixed-flavor-number scheme (FFNS) developed in
\cite{Nasson}-\cite{Bojak}.
In this approach the number of active flavors in the initial state
is limited to $n_f=4$, e.g., $u({\bar u}),d({\bar d}),s({\bar s})$
and $c({\bar c})$ quarks being the initial partons, whereas the
$b({\bar b})$ quark appears only in the final state. In this case
the beauty quark is always treated as a heavy particle, not as a
parton. In this scheme the mass of heavy quarks acts as a cutoff
parameter for the initial-state and final-state collinear singularities
and sets the scale for perturbative calculations. Actually, the
FFNS with $n_f=4$ is limited to a rather small range of transverse
momenta $p_t$ of produced $D$ or $B$ mesons less than the masses
of $c$ or $b$ quarks. In this scheme the terms $m_{c,b}^2/p_t^2$ 
are fully included.

Another approach is the so-called zero-mass
variable-flavor-number scheme (ZM-VFNS), see  
\cite{Kniehl1}-\cite{Greco} and references therein. 
It is the conventional parton model approach, the
zero-mass parton approximation  is also applied  to the $b$ quark,
although its mass is certainly much larger than the asymptotic
scale parameter $\Lambda_{QCD}$. In this approach the $b({\bar
b})$  quark is treated as an incoming parton originating from
colliding hadrons. 
This approach can 
be used in the region of large transverse momenta of produced charmed 
or beauty mesons, e.g., at $p_t\ge m_{c,b}$. Within this scheme the
terms of order $m_{c,b}^2/p_t^2$ can be neglected. Recently the
experimental inclusive $p_t$ spectra of $B$ mesons in $p{\bar p}$
collisions obtained by the CDFII Collaboration \cite{CDF1,CDF2} at
the Tevatron energy $\sqrt s=1.96$ TeV in the rapidity region
$-1\le y\le 1$ have been described rather satisfactorily within this
ZM-VFNS approach in \cite{Kniehl3} at $p_t\ge 10$ GeV/c using
the non-perturbative structure functions. In another kinematic
region, e.g., at $2.5$ GeV/c $\le p_t\le 10$ GeV/c, the FFNS model
allowed the CDFII data to be described without using
fragmentation functions of $b$ quarks to $B$ mesons. Both these
schemes have some uncertainties related to the renormalization
parameters. 

In this paper we study the charmed and beauty meson production
within the QGSM \cite{kaid1} or the dual parton model (DPM) \cite{Capella:1994}.
 based on the $1/N$ expansion in QCD \cite{tHooft:1974,Veneziano:1974}.  
We show that this approach can be applied rather successfully at not very large values
of $p_t$.
  
\section{General formalism}
\begin{figure}[ht]
\onefigure[height=2.cm,width=8.cm]{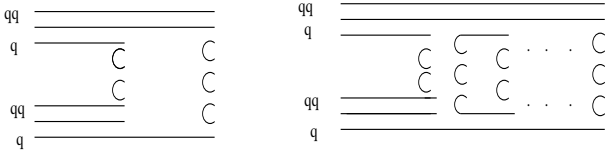} \caption{The
one-cylinder graph (left diagram) and the multi-cylinder graph
(right diagram) for the inclusive $p p\rightarrow h X$ process.}
\label{fig1}
\end{figure}

 Let us analyze the $D$-meson production in the $pp$ and
$p{\bar p}$ collisions within the QGSM 
including the transverse motion of quarks and diquarks in
colliding protons \cite{LAS}. As is known, the cylinder type
graphs for the $pp$ collision presented in fig.~\ref{fig1} make
the main contribution to this process \cite{kaid1}. 
The left diagram of fig.~\ref{fig1}, the so-called
one-cylinder graph, corresponds to the case where two colorless
strings are formed between the quark/diquark ($q/qq$) and the
diquark/quark ($qq/q$) in colliding protons; then, after their
breakup, $q{\bar q}$ pairs are created and fragmentated to a hadron,
for example, $D$ meson. The right diagram of fig.~\ref{fig1}, the
so-called multicylinder graph, corresponds to creation of the same
two colorless strings and many strings between sea
quarks/antiquarks $q/{\bar q}$ and sea antiquarks/quarks ${\bar
q}/q$ in the colliding protons.

The general form for the invariant inclusive hadron spectrum
within the QGSM is \cite{kaid2,LAS}

\begin{eqnarray}
E\frac{\upd\sigma}{\upd^3{\bf p}}\equiv
\frac{2E^*}{\pi\sqrt{s}}\frac{\upd\sigma}{\upd x\upd p_t^2}=
\sum_{n=1}^\infty \sigma_n(s)\phi_n(x,p_t)~, \label{def:invsp}
\end{eqnarray}

where $E,{\bf p}$ are the energy and the three-momentum of the
produced hadron $h$ in the laboratory system (l.s.) of colliding protons, 
$E^*,s$ are the energy of $h$ and the square of the initial energy in the
c.m.s of $pp$, $x,p_t$ are the Feynman variable and the transverse
momentum of $h$; $\sigma_n$ is the cross section for production of
the $n$-Pomeron chain (or $2n$ quark-antiquark strings) decaying
into hadrons, calculated within the ``eikonal approximation''
\cite{Ter-Mart}, the function $\phi_n(x,p_t)$ has the following
form \cite{LAS}:

\begin{eqnarray}
\phi_n(x,p_t)=\int_{x^+}^1 \upd x_1\int_{x_-}^1 \upd
x_2\psi_n(x,p_t;x_1,x_2)~, \label{def:phin}
\end{eqnarray}
where

\begin{eqnarray}
\psi_n(x,p_t;x_1,x_2) = \\ \nonumber
F_{qq}^{(n)}(x_+,p_t;x_1)F_{q_v}^{(n)}(x_-,p_t;x_2)/F_{q_v}^{(n)}(0,p_t) 
\\ \nonumber
+F_{q_v}^{(n)}(x_+,p_t;x_1)F_{qq}^{(n)}(x_-,p_t;x_2)/F_{qq}^{(n)}(0,p_t) 
\\ \nonumber
+2(n-1)F_{q_s}^{(n)}(x_+,p_t;x_1)F_{{\bar q}_s}^{(n)}(x_-,p_t;x_2)
/F_{q_s}^{(n)}(0,p_t). \label{def:psin}
\end{eqnarray}

and $x_{\pm}=0.5(\sqrt{x^2+x_t^2}\pm x),
x_t=2\sqrt{(m_h^2+p_t^2)/s}$,

\begin{eqnarray}
F_\tau^{(n)}(x_\pm,p_t;x_{1,2}) = \\ \nonumber
\int d^2k_t{\tilde f}_\tau^{(n)}(x_{1,2},k_t){\tilde G}_{\tau\rightarrow h}
\left(\frac{x_\pm}{x_{1,2}},k_t;p_t)\right),
\label{def:Ftaux}
\end{eqnarray}

\begin{eqnarray}
F_\tau^{(n)}(0,p_t) = \\ \nonumber
\int_0^1 dx^\prime d^2k_t {\tilde
f}_\tau^{(n)}(x^\prime,k_t) {\tilde G}_{\tau\rightarrow
h}(0,p_t)={\tilde G}_{\tau\rightarrow h}(0,p_t)~.
\label{def:Ftauzero}
\end{eqnarray}

Here $\tau$ means the flavor of the valence (or sea) quark or
diquark, ${\tilde f}_\tau^{(n)}(x^\prime,k_t)$ is the quark
distribution function depending on the longitudinal momentum
fraction $x^\prime$ and the transverse momentum $k_t$ in the
$n$-Pomeron chain; ${\tilde G}_{\tau\rightarrow h}(z,k_t;p_t)=
z{\tilde D}_{\tau\rightarrow h}(z,k_t;p_t)$, ${\tilde
D}_{\tau\rightarrow h}(z,k_t;p_t)$ is the fragmentation function
(FF) of a quark (antiquark) or diquark of flavor $\tau$ into a hadron
$h$ ($D$ meson in our case). 
We present the quark distributions
 and the FF in the factorized forms ${\tilde f}_\tau(x,k_t)=
f_\tau(x)g_\tau(k_t)$, and, according to \cite{LS:1992},
${\tilde G}_{\tau\rightarrow h}(z,k_t;p_t)=G_{\tau\rightarrow h}(z)
{\tilde g}_{\tau\rightarrow h}({\tilde k}_t)$, where ${\tilde{\bf k}}_t=
{\bf p}_t-z{\bf k}_t$.
 We take the quark distributions $f_\tau(x)$ and the FF  $G_{\tau\rightarrow h}(z)$
obtained within the QGSM from \cite{kaid2,Shabelsky,Piskunova},
whereas their $k_t$ distributions are chosen in the form suggested in
\cite{Piskunova:1984,LS:1992}

\begin{eqnarray}
g_\tau(k_t)=(B_q^2/2\pi)\exp(-B_q(\sqrt{k^2_t+m^2_D}-m_D))~,
\label{def:ktdistr}
\end{eqnarray}

\begin{eqnarray}
{\tilde g}_{\tau\rightarrow h}({\tilde k}_t)=(B_c^2/2\pi)
\exp(-B_c(\sqrt{{\tilde k}^2_t+m^2_D}-m_D)).
\label{def:ktFF}
\end{eqnarray}

After the integration of eq.~(\ref{def:Ftaux}) over $\upd^2k_t$ 
we have, according to \cite{LS:1992}, 

\begin{eqnarray}
F_\tau^{(n)}(x_\pm,p_t;x_{1,2})={\tilde f}_\tau^{(n)}(x_{1,2})
G_{\tau\rightarrow h}(z)I_n(z,p_t)~, 
\label{def:Fnew}
\end{eqnarray}

where $z=x_\pm/x_{1,2}$,
$I_n(z,p_t)=B_z^2/(2\pi(1+B_zm_D))\exp(-B_z(m_{Dt}-m_D))$, 
$m^2_{Dt}=p_t^2+m_D^2$; $B_z=B_c/(1+n\rho z^2)$, $\rho=B_c/B_q$. 
The function $B_z$ also can be presented in the equivalent form
$B_z=B_q/({\tilde\rho}+n z^2)$, where ${\tilde\rho}=B_q/B_c$. 
The differential cross section $\upd\sigma/\upd p_t^2$
for $D$ mesons produced in $pp$ collisions is written in the  
form \cite{kaid2}

\begin{eqnarray}
\frac{\upd\sigma}{\upd p_t^2}=\frac{\pi}{2}
\sqrt{s}\sum_{n=1}^\infty\sigma_n(s) \int\phi_n(x,p_t)\frac{\upd
x}{E^*}~. \label{def:dsigdptsq}
\end{eqnarray}

\begin{figure}[ht]
\onefigure[scale=0.55]{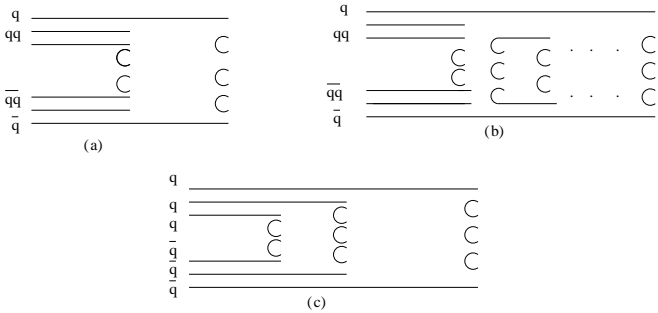}
 \caption{The one-cylinder graph (left diagram) and
 the multi-cylinder graph (right diagram),
 and the three-chains graph (down diagram) for the
 $p{\bar p}\rightarrow h X$ inclusive process.} 
\label{fig2}
 \end{figure}

 The production of heavy
mesons like $D$- and $B$-mesons in proton-antiproton collisions at
high energies is usually analyzed within the different schemes of
QCD. To study these processes within the QGSM we have to include
at least one additional graph corresponding to the creation of
three chains between quarks in the initial proton and antiquarks
in the colliding antiproton, as is illustrated in fig.~\ref{fig2}
(bottom diagram).

The left and right diagrams in fig.~\ref{fig2} are similar to the
one-cylinder and multicylinder diagrams for the $pp$ collision in
fig.~\ref{fig1} with a following difference. In the $p{\bar p}$
collision two colorless strings between quark/diquark ($q/qq$) in
the initial proton and antiquark/antidiquark (${\bar q}/{\bar
qq}$) are created. Many quark-antiquark ($q-{\bar q}$) strings for
$p{\bar p}$ collision (fig.~\ref{fig2}, right diagram) are the
same as for the $pp$ collision (fig.~\ref{fig1}, right diagram).
Therefore, the invariant inclusive spectrum of hadrons produced in
the $p{\bar p}$ collision calculated within the QGSM has the
following form:

\begin{eqnarray}
E\frac{\upd\sigma^{p{\bar p}}}{\upd^3{\bf p}}=
\sigma_1(s)((1-\omega)\phi^{p{\bar p}}_1(x,p_t)+\omega{\tilde\phi}(x,p_t))+\\
\nonumber
+\sum_{n=2}^\infty \sigma_n(s)\phi^{p{\bar p}}_n(x,p_t)
\label{def:sppbarp}
\end{eqnarray}

where $1-\omega$ is the probability of contribution of the cut
one-cylinder (one-Pomeron exchange) and cut multicylinder
(multiPomeron exchanges) graphs (the left and right
diagrams in fig.~\ref{fig2}), whereas $\omega$ is the probability
of the contribution of the three-chain diagram 
to the inclusive spectrum. The value of $\omega$ can be estimated
as the ratio of the $p{\bar p}$ annihilation cross section
$\sigma_{p{\bar p}}^{ann}$ to the total $p{\bar p}$ cross section
$\sigma_{p{\bar p}}^{tot}$. The cross section $\sigma_{p{\bar
p}}^{tot}$ is well known in the wide range of the initial energies
to the Tevatron energy, whereas experimental data on
$\sigma_{p{\bar p}}^{ann}$ are available only for the antiproton
initial energy about $10$ GeV, see \cite{Uzh02} and references
therein. However, some theoretical predictions, for example
\cite{GosNus80, BZK:1988}, show that asymptotically
$\sigma_{p{\bar p}}^{ann}$ goes to about $2-4$ mb. It corresponds
to $\omega\simeq \sigma_{p{\bar p}}^{ann}/\sigma_{p{\bar
p}}^{tot}~<~ 0.1$ at the Tevatron energy. Note that in addition
to the graph of fig~\ref{fig2}c there can be diagrams consisting of these three chains 
and multicilynder chains between sea quarks and antiquarks. However, as our estimations 
show, their contribution to the inclusive spectrum is much smaller than the contribution
from the three-chain graph (fig~\ref{fig2}c). Therefore we neglect 
it. 
The form for the function $\phi^{p{\bar p}}_n(x,p_t)$ is similar
to $\phi_n(x,p_t)$ entering into (3) by replacing
$F_{q_v}^{(n)}(x_-,p_t;x_2)$, $F_{q_v}^{(n)}(0,p_t)$ to $F_{\bar
{qq}}^{(n)}(x_-,p_t;x_2)$, $F_{\bar {qq}}^{(n)}(0,p_t)$
respectively, and replacing $F_{qq}^{(n)}(x_-,p_t;x_2)$,
$F_{qq}^{(n)}(0,p_t)$ to $F_{{\bar q}}^{(n)}(x_-,p_t;x_2)$ and
$F_{{\bar q}}^{(n)}(0,p_t)$ respectively. The additional term
${\tilde\phi}(x,p_t)$ in (\ref{def:sppbarp}) has the
following form

\begin{eqnarray}
{\tilde\phi}(x,p_t)=3{\tilde F}_{q_v}(x_+,p_t){\tilde F}_{{\bar
q}_v}(x_-,p_t)/{\tilde F}_q(0,p_t)~, \label{def:tphi}
\end{eqnarray}

where ${\tilde F}_{q_v({\bar
q}_v)}(x_\pm,p_t)=F^{(n=1)}_{q_v({\bar q}_v)}(x_\pm,p_t)$ and
${\tilde F}_q(0,p_t)=F_q^{(n=1)}(0,p-t)$.

\section{Results and discussion}
\begin{figure}
\onefigure[scale=1]{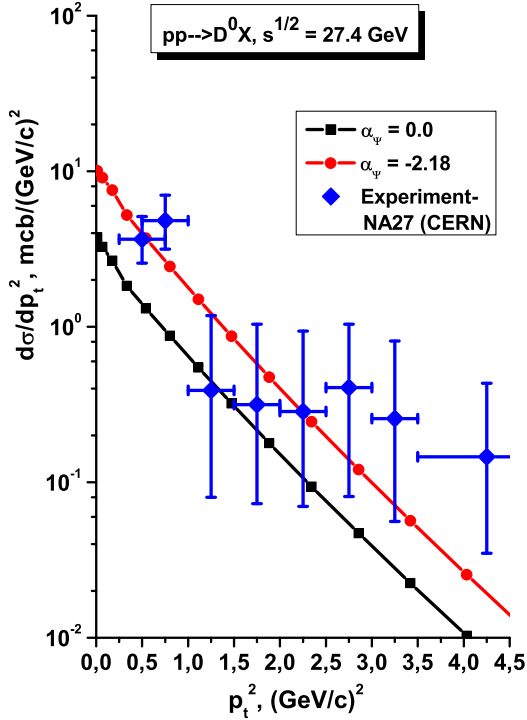} \caption{The inclusive
spectrum for $D^0$ mesons produced in the $pp$ collision at
$\sqrt{s}=27.4$ GeV as a function of $p_t^2$ , see details
in \cite{LS:1992}.} 
\label{fig3}
\end{figure}

\begin{figure}
\onefigure[scale=1]{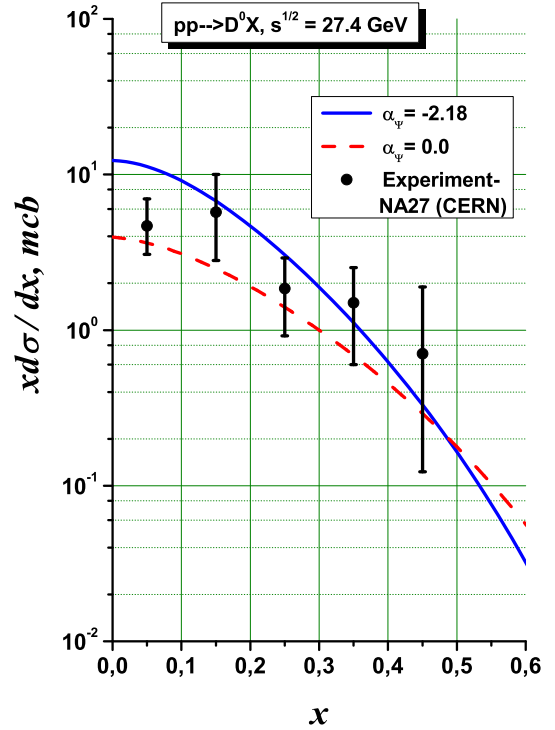} \caption{The inclusive
spectrum for $D^0$ mesons produced in the $pp$ collision at
$\sqrt{s}=27.4$ GeV as a function of $x$, see details
in \cite{LS:1992}.}
\label{fig4}
\end{figure}

To illustrate our approach we present in
figs.~\ref{fig3},\ref{fig4} the inclusive spectrum
$d\sigma/dp_t^2$ of $D^0$-mesons produced in the reaction
$pp\rightarrow D^0 X$ at $\sqrt{s}=27.4$ GeV as a function of
$p_t^2$ given by eq.(\ref{def:dsigdptsq}).  
One can see from fig.~\ref{fig3} that the use of different
values for the intercept $\alpha_\Psi(0)$ leads mainly to the
shift of the theoretical lines along the $y$ axis. 
figs.~\ref{fig3},\ref{fig4} show
a satisfactory description of the experimental data obtained 
by the NA27 Collaboration \cite{NA27} at both values
for the linear $\Psi$-trajectory when $\alpha_\Psi(0)=-2.18$ and
for the nonlinear one when $\alpha_\Psi(0)=0$. Unfortunately, the
experimental data are very poor because of big error bars;
therefore, we cannot get new information on the $\Psi$-trajectory.

\begin{figure}
\onefigure{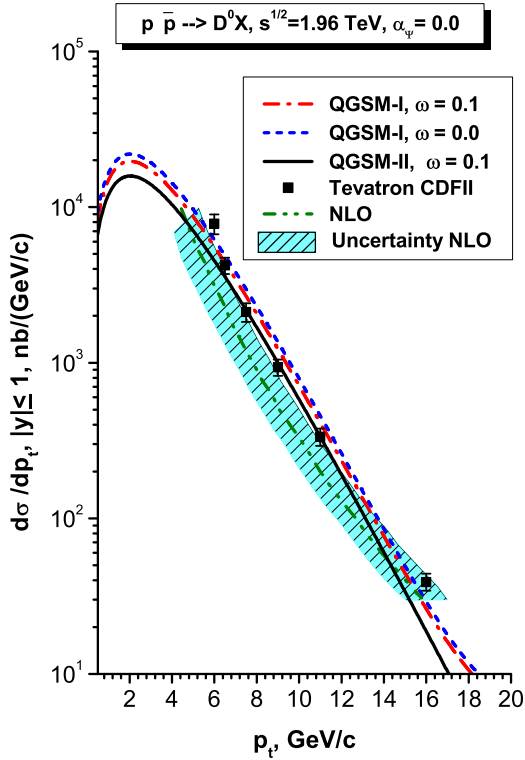} 
\caption{The inclusive $p_t$-spectrum
for $D^0$ mesons produced in the $p{\bar p}$ collision at the
Tevatron energy $\sqrt{s}=1.96$ TeV obtained within the QGSM
(the solid and dashed lines) and within the NLO QCD
\cite{CDF_Dmes} (the hatched regions); QGSM-I:
$B_q\simeq 0.18$(Gev/c)$^{-1}$, $\tilde{\rho}\simeq 0.32$;
QGSM-II:
$B_q=4.5$(GeV/c)$^{-1}$, ${\tilde\rho}=7$.} 
\label{fig5}
\end{figure}

\begin{figure}
\onefigure{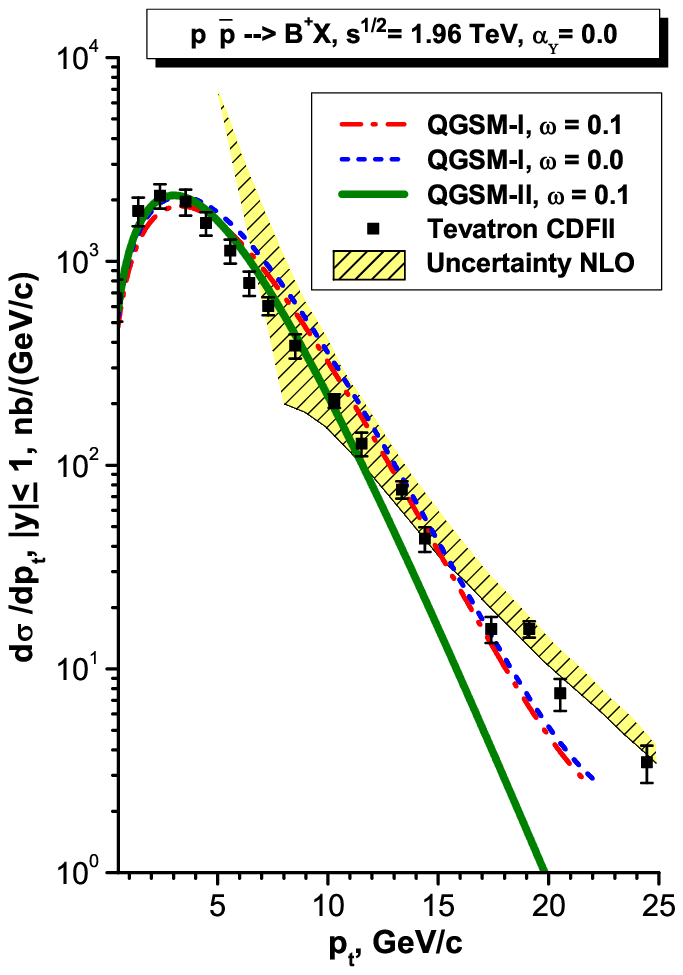} 
\caption{The inclusive $p_t$-spectrum
for $B^+$ mesons produced in the $p{\bar p}$ collision at the
Tevatron energy $\sqrt{s}=1.96$ TeV obtained within the QGSM
(the solid and dashed lines) and within the NLO QCD
\cite{CDF_Dmes} (the hatched regions); QGSM-I: 
$B_q\simeq 0.21$(GeV/c)$^{-1}$, $\tilde{\rho}\simeq 0.32$; 
QGSM-II is the same as in Fig~(\ref{fig5}).}    
\label{fig6}
\end{figure}

\begin{figure}[ht]
\onefigure[height=8.5cm,width=8.5cm]{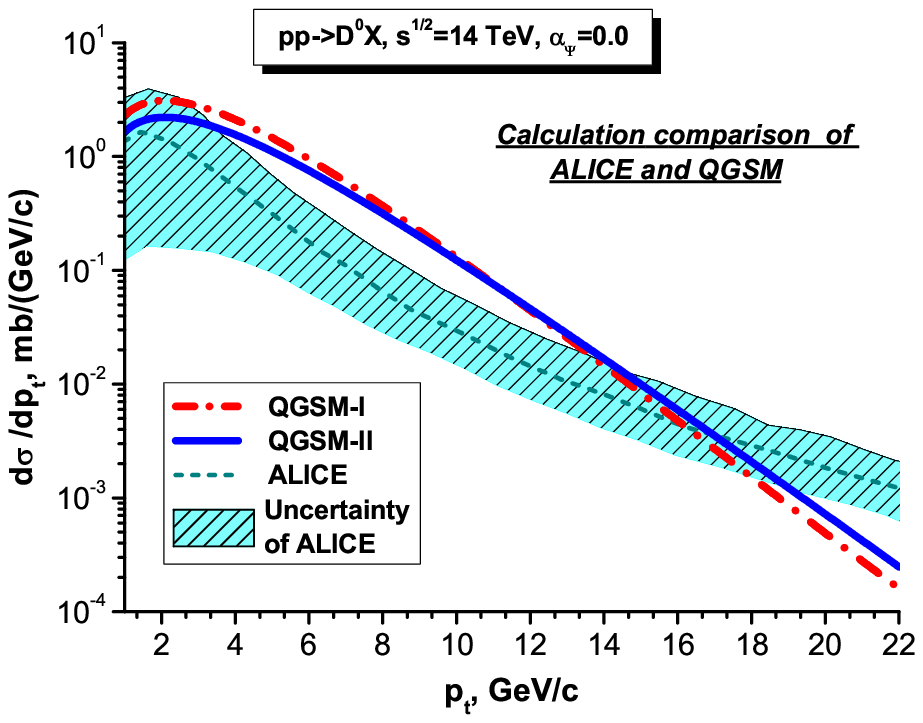}
\caption{The inclusive spectrum for $D^0$ mesons produced in the
$pp$ collision at the LHC energy $\sqrt{s}=14$ TeV obtained within
the QGSM for charmed mesons and the NLO QCD for $c$ quarks
 \cite{ALICE}; QGSM-I corresponds to the same QGSM-I 
as in Fig~\ref{fig5};
QGSM-II: 
$B_q=4.$ (GeV/c)$^{-1}$, $\tilde{\rho}=7$.} 
\label{fig7}
\end{figure}

\begin{figure}[ht]
\onefigure[height=8.5cm,width=8.5cm]{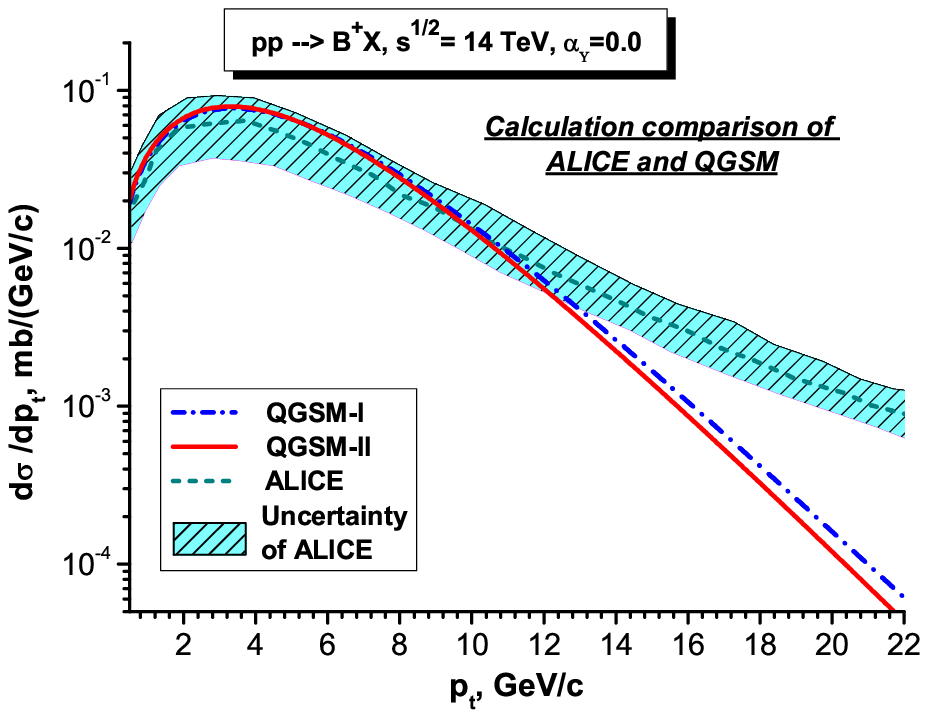}
\caption{The inclusive spectrum for $B^+$ mesons produced in the
$pp$ collision at the LHC energy $\sqrt{s}=14$ TeV obtained within
the QGSM for beauty mesons and the NLO QCD for single
$b$ quarks \cite{ALICE}; QGSM-I corresponds to the same QGSM-I
as in Fig~(\ref{fig6});
QGSM-II is the same as in Fig~\ref{fig7}.}  
\label{fig8}
\end{figure}

The inclusive $p_t$ spectra of $D^0$ and $B^+$ mesons produced in
the $p{\bar p}$ collision at the Tevatron energy $\sqrt{s}=1.96$
TeV are presented in figs.~\ref{fig5},\ref{fig6}. 
The hatched regions in
figs.~\ref{fig5},\ref{fig6} show the calculations within the NLO
QCD including uncertainties \cite{CDF_Dmes}. 
The dashed lines ($\omega=0$) and 
dash-dotted curves ($\omega=0.1$) in  figs.~\ref{fig5},\ref{fig6} correspond to our best fit
obtained within the QGSM-I calculations using the parameter values $B_c=0.65$(GeV/c)$^{-1}$ 
for $D$ mesons (fig.~\ref{fig5}) and $B_c=0.55$(GeV/c)$^{-1}$ for $B$ mesons (fig.~\ref{fig6}), and 
$\rho\equiv B_c/B_q=3.1$ \cite{LS:1992}, e.g., $B_q\simeq 0.2$(GeV/c)$^{-1}$. 
However, the value used for the slope $B_q$ of the quark distribution as a function of $k_t$ is
too small. Therefore, we also calculated these $p_t$-spectra taking the more realistic
values $B_q=4.5$(GeV/c)$^{-1}$ at the Tevatron energy and $B_q=4$(GeV/c)$^{-1}$
at the LHC energy that correspond approximately to $<k_t>\simeq 0.45$ GeV/c and  $<k_t>\simeq 0.5$ GeV/c
respectively. This calculation (QGSM-II) is shown by the solid lines in
figs.~\ref{fig5},\ref{fig6} and in figs.~\ref{fig7},\ref{fig8}. According to the experimental data,
the mean transverse momentum of hadrons produced in hadron collisions slowly increases
as the energy increases. Therefore, the internal transverse momentum of the quark in
the proton can also slow increase.     
In fact, we have only one free parameter $\tilde{\rho}\equiv 1/\rho=B_q/B_c$ which is experimentally  
unknown, whereas $B_q$ is directly related to the mean transverse momentum of the quark in the proton or 
antiproton which is more or less known experimentally.
To describe the experimental 
data on the $p_t$-spectra in the $p_t$ region where the NLO QCD calculation has a big uncertainty
we chose $\tilde{\rho}=7$ both for the Tevatron and the LHC energies.
When $B_c~<~1$ (GeV/c)$^{-1}$, eq.(\ref{def:ktFF}) can be approximately presented in the form 
${\tilde g}_{\tau\rightarrow h}({\tilde k}_t)\simeq a^2/(a^2+{\tilde k}^2_t)$ at
${\tilde k}^2_t<2m_D$, where $a=\sqrt{2m_D/B_c}$. 
This form is similar to the form for the FF of heavy quarks obtained within the perturbative 
QCD, see for example \cite{Greco:1994}.
Note that the function $I_n(z,p_t)$ in eq.(\ref{def:Fnew}) was obtained in 
\cite{LS:1992,LAS} on the assumption of the consequent sharing of the transverse  momentum
$p_t$ in the proton (antiproton) between $n$-Pomeron chains. It allowed us to describe rather 
satisfactorily the experimental data on the inclusive $p_t$ spectra of charmed and beauty mesons
produced in $p{\bar p}$ collisions at moderate values of the transverse momentum $p_t~<~10$ GeV/c.
It is illustrated in figs.~\ref{fig5},\ref{fig6}. We also use  
$\lambda=2\alpha_{D^*(B^*)}^\prime(0)<p^2_t>$,
$\alpha_{D^*(B^*)}^\prime(0)\simeq 0.5$ (GeV/c)$^{-2}$ is the
slope of the $D^*$ or $B^*$ Regge trajectory, $<p^2_t>\simeq 5$ 
(GeV/c)$^2$ is the mean transverse momentum squared of the $D$ meson 
or $B$ meson that was found from the CDFII experimental data.
Note that the our calculation showed that the contribution
of the three-chain graph (fig.~\ref{fig2}c) is very small at the
Tevatron energy. It is due to small values of the $p{\bar p}$ annihilation cross 
section at very high energies \cite{GosNus80,BZK:1988}.   

The predictions for inclusive $p_t$ spectra of $D^0$ and
$B^+$ mesons produced in the $pp$ collision at LHC energies and
the NLO QCD calculation for the produced charmed quarks
\cite{ALICE} are presented in figs.~\ref{fig7},\ref{fig8}.

The solid  lines correspond to our calculations within the
{QGSM-II}, whereas the hatched regions show the calculations within the
NLO QCD including uncertainties \cite{ALICE}. A big difference
between the QGSM and NLO QCD calculations at $p_t>10$
GeV/c for $D$ and $B$ mesons can be due to the following.
First, the NLO QCD calculation \cite{ALICE}
does not include the hadronization of quarks to heavy mesons,
whereas the QGSM calculation includes it. Second, we do not
include the contribution of gluons and their hard scatterings off
quarks and gluons which can be sizable at large values of $p_t$.

\section{Conclusion}
 We have shown that the modified QGSM including the
intrinsic longitudinal and transverse motion of quarks
(antiquarks) and diquarks in colliding protons allowed us to
describe rather satisfactorily the existing experimental data on
inclusive spectra of $D$ mesons produced in $pp$ collisions and
to make some predictions for similar spectra at LHC energies. To
verify whether these predictions can be reliable or not we apply
the QGSM to the analysis of charmed and beauty meson production in
proton-antiproton collisions at Tevatron energies including graphs
like those in fig.~\ref{fig2}c corresponding to annihilation of
quarks and antiquarks in colliding $p$ and ${\bar p}$, and
production of $D$-mesons.

We got a satisfactory QGSM-II description ($p_t~<~10$ GeV/c)
of the experimental data on
$p_t$ spectra of $D^0$ and $B^+$ mesons produced in the $p{\bar
p}$ collisions which were obtained by the CDFII Collaboration at
the Tevatron \cite{CDF_Dmes}. 
At larger values of $p_t$ the calculations within the NLO of QCD 
\cite{Nasson,Kniehl1} result in a better description of these data.
It can be due to the contribution of gluons inside the colliding 
proton and antiproton which can interact with other gluons and quarks
(antiquarks) and fragmentate to charmed mesons. This effect is not 
taken into account in the presented QGSM. Therefore, the QGSM including the internal
transverse momenta of partons in the proton (let us call it the ``soft QCD'')
allows us to describe the inclusive $p_t$ spectra of heavy mesons produced 
in $pp$ and $p{\bar p}$ collisions at high energies at not very large values
of $p_t$. 
To describe these spectra and make some predictions for the future LHC experiments
in a wide region of transverse momenta one can combine the ``soft QCD'' at small 
values of $p_t$ with the NLO QCD at large $p_t$.

We found that the $p_t$ spectra of $D$ and $B$ mesons calculated within 
the QGSM are almost insensitive to the form of the sea $c({\bar c})$ and
$b({\bar b})$ quark distributions in colliding
protons/antiprotons. To find a new information on it we intend to
study the charm and beauty hadron production in $pp$ collisions at
small scattering angles.

\acknowledgments We thank W. Cassing, M. Deile, A. V. Efremov, K. Eggert,
D. Elia, S.B.Gerasimov, A. B. Kaidalov, B. Z. Kopeliovich, A. D. Martin, 
M. Poghosyan, K. Safarik, J. Schukraft, V. V. Uzhinsky and 
D. Weber for very useful discussions. This work was supported in part by the 
High Energy Foundation and the World Science Agency and the RFBR grant N 08-02-01003.

\end{document}